\documentclass[11pt]{article} 
\usepackage{mystyle-new}
\usepackage{authblk}
\usepackage[T1]{fontenc}
\usepackage{epsfig,amsmath} 
\usepackage{hepnames,hepunits}
\usepackage{hyperref}
\usepackage{color}
\usepackage{graphicx}
\usepackage{orcidlink}
\definecolor{red}{rgb}{1,0,0}
\def\lesssim{\ \hbox{\raise 2pt \hbox{$<$} \kern -13pt
                     \lower 3pt \hbox{$\sim$}}\ }
\def\greatersim{\ \hbox{\raise 2pt \hbox{$>$} \kern -13pt
                     \lower 3pt \hbox{$\sim$}}\ }

\def\lsim{\mathrel{\rlap{\lower4pt\hbox{\hskip1pt$\sim$}}
    \raise1pt\hbox{$<$}}}                
\def\gsim{\mathrel{\rlap{\lower4pt\hbox{\hskip1pt$\sim$}}
    \raise1pt\hbox{$>$}}}                

\def\cascade{{\sc Cascade}}
\def\pythia{{\sc Pythia}}

\def\mcatnlo{{MCatNLO}}

\input epsf.tex
\def\desepsf(#1 width #2){\epsfxsize=#2 \epsfbox{#1}}
\def\kt{\ensuremath{k_{\rm T}}}

\def\pt{\ensuremath{p_{\rm T}}}
\def\PZ{\ensuremath{Z}}

\newcommand{\as}{\ensuremath{\alpha_s}}

\newcommand{\PBM}{PB}

\newcommand{\MCatNLO}{{\sc MadGraph5\_aMC@NLO}}
\newcommand{\CAS}{{\mcatnlo+CAS3}}
\newcommand{\jpsi}{{\ensuremath{J/\psi}}}

\newcommand{\GeV}{GeV}
\newcommand{\TeV}{TeV}

\newenvironment{tolerant}[1]{\par\tolerance=#1\relax}{ \par }
\usepackage{amsmath,bm}
\usepackage{lineno}

\usepackage{cite,./mcite}
\usepackage{tikz}
\usepackage[symbol]{footmisc}

\providecommand{\DOI}[1]{\href{http://dx.doi.org/#1}}

\begin{document}

\title{Comparison of CMS measurements with predictions at NLO applying the Parton Branching Method and \pythia }
\author[1]{Fernando~Guzman~\orcidlink{0000-0002-7612-1488}}
\author[1]{Si~Hyun~Jeon~\orcidlink{0000-0003-1208-6940}}
\author[1]{Hannes~Jung~\orcidlink{0000-0002-2964-9845}}
\author[1]{Danyer~Perez~Adan~\orcidlink{0000-0003-3416-0726}}
\author[1]{Sara~Taheri~Monfared~\orcidlink{0000-0003-2988-7859}}
\affil[1]{Supervisor}
\author[2]{Fateme~Almaksusi~\orcidlink{0009-0002-5932-7967}}
\author[2]{Daniel~Belmonte~Perez~\orcidlink{0009-0000-6372-4709}}
\author[2]{Dorukhan~Boncukcu~\orcidlink{0000-0003-0393-5605}}
\author[2]{Aleksandr~Boger~\orcidlink{0009-0005-6584-687X}}
\author[2]{Emmanuel~Botero~Osorio~\orcidlink{0009-0000-7599-7640}}
\author[2]{Isadora~Bozza~Galv\~ao~\orcidlink{0000-0001-7707-2852}}
\author[2]{Juan~Esteban~Ospina~Holguin~\orcidlink{0009-0002-8559-6386}}
\author[2]{Behnam~Falahi~\orcidlink{0009-0007-0923-3205}}
\author[2]{Faeze~Gagonani~\orcidlink{0009-0000-6218-6320}}
\author[2]{Omar~Gonzalez~\orcidlink{0009-0004-3221-5134}}
\author[2]{Acelya~Deniz~G\"ung\"ord\"u~\orcidlink{0009-0005-0460-0980}}
\author[2]{Abdelhamid~Haddad~\orcidlink{0000-0001-9553-9372}}
\author[2]{Mahtab~Jalalvandi~\orcidlink{0009-0000-9277-1555}}
\author[2]{Josue~Daniel~Jaramillo~\orcidlink{0009-0000-4320-7595}}
\author[2]{Jesus~Jimenez~Zepeda~\orcidlink{0009-0002-3390-8085}}
\author[2]{Gleb~Kutyrev~\orcidlink{0009-0001-5621-8629}}
\author[2]{Nazanin~Zahra~Noroozi~\orcidlink{0009-0007-7452-3233}}
\author[2]{Nestor~Raul~Mancilla~Xinto~\orcidlink{0000-0001-5968-2710}}
\author[2]{Haritz~Mentaste~\orcidlink{0009-0009-5432-1671}}
\author[2]{Lucas~Johnny~Monte~Tamayo~\orcidlink{0009-0005-0266-901X}}
\author[2]{Fernanda~Mora~Rey~\orcidlink{0000-0002-7780-7908}}
\author[2]{Inmaculada~Moyano-Rejano~\orcidlink{0009-0005-1664-0760}}
\author[2]{Vibha~Padmanabhan~\orcidlink{0000-0003-4448-8258}}
\author[2]{Mayvi~Pedraza~\orcidlink{0009-0000-4289-7790}}
\author[2]{Swapnil~Rathore\orcidlink{0009-0009-8251-9765}}
\author[2]{Cristina~Ruiz~Gonzalez~\orcidlink{0009-0001-0797-8445}}
\author[2]{Caue~E.~Sousa~\orcidlink{0000-0001-5380-8327}}
\author[2]{Quratulain~Zahoor~\orcidlink{0009-0001-7427-4368}}
\author[2]{Haoliang~Zheng~\orcidlink{0009-0002-4032-1493}}
\affil[2]{Summerstudent}

\date{}
\begin{titlepage} 
\maketitle

\begin{abstract}
In August 2023, more than 30 students joined the {\it Special Remote DESY summer-school} to work on projects of importance for LHC experiments. In a dedicated initiative, analyses that had not been incorporated into the RIVET package were implemented and verified. Here, a brief description of the accomplished work is given, and a comparison of the measurements with predictions obtained from matched standard parton shower Monte Carlo event generators as well as with those obtained from Parton-Branching TMDs with corresponding parton showers are presented.
\end{abstract} 

\end{titlepage}

\section{Introduction}
In August 2023, more than 30 students participated the {\it Special Remote DESY summer-school}~\cite{Summerschool2023} to work on projects of importance for LHC experiments. The {\it Special Remote DESY summer-school} was endorsed by IUPAP~\cite{IUPAP-Summerschool2023} and included in the program of the International Year of Basic Sciences for Sustainable Development (IYBSSD)~\cite{IYBSSD-Summerschool2023}.  The school was fully online, with virtual meeting sessions from 2-4 pm CEST, for students from the far East it was late in the evening and for students from the far West it was very early in the morning. A similar fully online school was held in 2021 \cite{Abdulhamid:2021dpr}.

The school in 2023 was held in the spirit of the IUPAP policy to {\it "open the channels for scientific cooperation across all political and other divides in the hope and expectation that enhanced scientific collaborations are an important means to develop improved understandings between different peoples that contributes to world peace"}.

The student projects were on coding, testing and validating computer codes of experimental analyses which have been already published by the CMS collaboration, but are not yet available in the Rivet (Robust Independent Validation of Experiment and Theory) package \cite{Bierlich:2019rhm}. 

In the following, we describe briefly the analyses in the area of QCD jets and electro-weak physics, which have been developed. 

\section{New Rivet plugins}
Several published results obtained with the CMS experiment at the LHC, which were not yet implemented in the Rivet package were investigated, ranging form jet production at $\sqrt{s}=7$~\TeV , over heavy flavor production to measurements involving the electroweak (EW) bosons $\gamma, \PW ,\PZ $ at $\sqrt{s}=13$~\TeV .
The theoretical predictions at leading order (LO), obtained with \pythia 8~\cite{Sjostrand:2007gs,Sjostrand:2014zea,Bierlich:2022pfr}, using the CUETM1 tune~\cite{Khachatryan:2015pea} to fix a number of free parameters sensitive to the treatment of the underlying event, were used to validate the implementation of the different analyses into the Rivet package, and to compare the obtained results with those shown in the original publications.

The following analyses were studied:
\begin{tolerant}{2000}
\begin{itemize}
\item {\tt CMS\_2010\_I878118} Prompt and Non-Prompt $J/\psi$ Production in pp Collisions at $\sqrt{s}=7$ TeV~\cite{CMS:2010nis} (Mayvi Pedraza Monzon, Supervisor:  F. Guzman)
\item {\tt CMS\_2011\_I944755} $J/\psi$ and $\psi_{2S}$ production in pp collisions at $\sqrt{s}=7$ TeV  \cite{CMS:2011rxs} (Shijie Zhang, Supervisor: H. Jung)
\item {\tt CMS\_2011\_I895742} Measurement of the differential dijet production cross section in proton-proton collisions at $\sqrt{s}=7$ TeV \cite{CMS:2011nlq} 
(Josu\'e Daniel Jaramillo Arroyave, Fernanda Mora Rey, Supervisor: H.~Jung)
\item {\tt CMS\_2012\_I1093951} Measurement of the Inclusive Cross Section $\sigma (pp \to b\bar{b} X \to \mu \mu X^{\prime}$ at $\sqrt{s}=7$ TeV \cite{CMS:2012xsp}
 (Faeze Gagonani, Nestor Raul Mancilla Xinto, Supervisor: H.~Jung)
\item {\tt CMS\_2015\_I1345159} Distributions of Topological Observables in Inclusive Three- and Four-Jet Events in pp Collisions at $\sqrt{s} = 7$ TeV \cite{CMS:2015lzt} 
(Dorukhan Boncukcu, Acelya~Deniz~G\"ung\"ord\"u, Supervisor: H.~Jung )
\item {\tt CMS\_2015\_I1359450} Measurement of the Z boson differential cross section in transverse momentum and rapidity in proton-proton collisions at 8 TeV \cite{Khachatryan:2015oaa}
(Emmanuel Botero Osorio, Jesus Jimenez-Zepeda, Nazanin Zahra Norouzi, Juan Esteban Ospina, Swapnil Rathore, Supervisor: S.~Taheri~Monfared)
\item {\tt CMS\_2015\_I1410826} Measurement of the inclusive jet cross section in pp collisions at $\sqrt{s} = 2.76\,\text {TeV}$ \cite{CMS:2015jdl}
(Vibha Padmanabha, Quratulain Zahoor, Supervisor: H.~Jung)
\item {\tt CMS\_2016\_I1485195} Measurement of the total and differential inclusive $B^+$ hadron cross sections in pp collisions at $\sqrt{s}$ = 13 TeV \cite{CMS:2016plw}
(Lucas Tamayo, Supervisor: H.~Jung)
\item {\tt CMS\_2019\_I1757506} Study of J/$\psi$ meson production inside jets in pp collisions at $\sqrt{s} =$ 8 TeV \cite{CMS:2019ebt}
(Behnam Falahi, Supervisor: H.~Jung)
\item {\tt CMS\_2021\_I1876550} Measurement of prompt open-charm production cross sections in proton-proton collisions at $ \sqrt{s} $ = 13 TeV \cite{CMS:2021lab}
(Fateme Almaksusi, Haoliang Zheng, Supervisor: H.~Jung)
\item {\tt CMS\_2021\_I1869513} Measurement of the electroweak production of Z$\gamma$ and two jets in proton-proton collisions at $\sqrt{s} =$ 13 TeV and constraints on anomalous quartic gauge couplings \cite{CMS:2021gme}
(Daniel Belmonte, Haritz Mentaste, Omar Gonz\'alez, Supervisor: Si~Hyun~Jeon)
\item {\tt CMS\_2021\_I1876311} Measurements of the electroweak diboson production cross sections in proton-proton collisions at $\sqrt{s} =$ 5.02 TeV using leptonic decays \cite{CMS:2021pqj}
(Cristina~Ruiz~Gonzalez, Inmaculada~Moyano-Rejano, Supervisor: Si~Hyun~Jeon)
\item{\tt CMS\_2021\_I1949191} Measurement of the inclusive and differential WZ production cross sections, polarization angles, and triple gauge couplings in pp collisions at $ \sqrt{s} $ = 13 TeV \cite{CMS:2021icx}
(Mahtab Jalalvandi, Supervisor: D. ~Perez~Adan)
\item {\tt CMS\_2021\_I1992937} Measurement of the production cross section for Z+b jets in proton-proton collisions at $\sqrt s$ = 13\,\,TeV \cite{CMS:2021pcj}
(Caue~E.~Sousa, Gleb Kutyrev, Aleksandr Boger, Isadora~Bozza~Galv\~ao, Abdelhamid Haddad, Supervisor: D.~Perez~Adan)
\end{itemize}
\end{tolerant}
\section{Comparison with predictions}

\begin{tolerant}{2000}
While predictions at leading order (LO) obtained with \pythia 8 were used to test and validate the coded analyses, we present here a comparison of theoretical predictions obtained at next-to-leading (NLO) in the strong coupling $\as$. 

At NLO, the \MCatNLO\ package~\cite{Alwall:2014hca} was used to calculate the hard process and supplement the partonic processes with parton shower and hadronization. We use the NLO Parton Branching (\PBM ) collinear and the transverse momentum dependent (TMD) 
 parton distributions as obtained in Ref. \cite{Martinez:2018jxt} from QCD fits to precise DIS (Deep Inelastic Scattering)  data from HERA~\cite{Abramowicz:2015mha} using the xFitter analysis framework \cite{xFitterDevelopersTeam:2022koz,Alekhin:2014irh}. We cross check the results with predictions using the NNPDF31 collinear parton distributions~\cite{Ball:2017nwa}.
We apply the Parton Branching transverse momentum distributions (TMD) PB-NLO-2018-Set 2 together with the TMD parton shower for the initial state as implemented in \cascade 3 \cite{Baranov:2021uol} (labelled \CAS ) as well as a standard  parton shower obtained with \pythia 8 (labelled \mcatnlo+\pythia 8). The NLO hard process is calculated 
with appropriate subtraction terms for the use with \cascade 3 and \pythia 8. We use $\mu_{def}$ as factorization and renormalization scale  $\mu_{def}=\mu_{R,F}=\frac{1}{2} \sum_i \sqrt{m^2_i +p^2 _{t,i}}$, where the index $i$ runs over all particles in the matrix element final state. 
\end{tolerant}

\begin{figure}[h!tb]
\begin{center}
\includegraphics[width=0.45\textwidth]{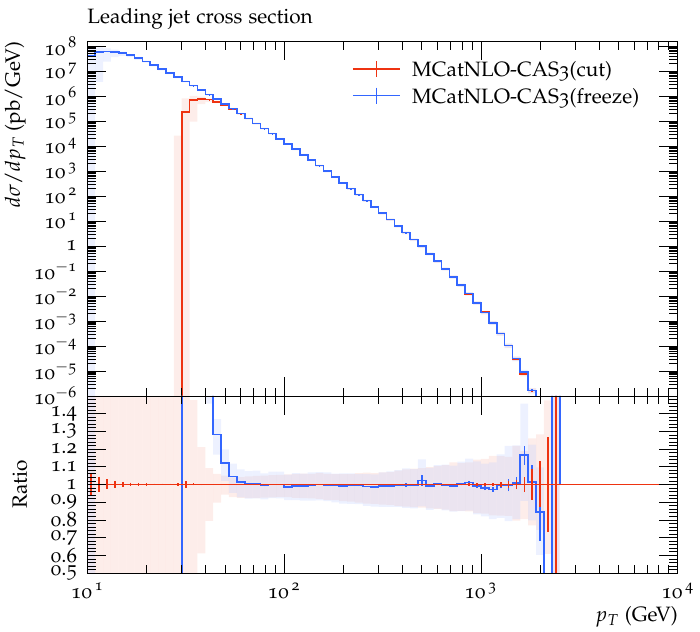}
  \caption{\small Differential cross section as a function of the leading jet $\pt$ obtained with  \protect\CAS\ applying to freeze of the scale (see eq.(\protect\ref{freeze})) and with a cutoff of $\pt> 30 $ \GeV.}
\label{dijets_CAS_set12}
\end{center}
\end{figure}

Since some measurements are performed at very small transverse momenta, special care has to be taken to avoid instabilities of the NLO cross section as well as to avoid too large negative cross sections. We apply a taming factor for  the scale $\mu_{R,F}$:
\begin{equation}
\mu=\mu_{R,F} = \mu_{def}^2 + \mu_0^2  \, ,
\label{freeze}
\end{equation}
with $\mu_{def}$ being the default scale  and $\mu_0$ being a free parameter to avoid too small scales. We found the $\mu_0 = 20 $ \GeV\ leads to stable results  for jet production with a transverse momentum down to $\pt \sim 10 $ \GeV .

In Fig.~\ref{dijets_CAS_set12} we show a  comparison of the cross section as a function of the leading jet $\pt$ calculated with and without the taming factor of eq.(\ref{freeze}). 
We do not observe any significant change in the cross section at larger \pt , however, applying eq.(\ref{freeze}) we are able to use an NLO calculation for jet production down to $\pt \sim 10 $ \GeV  .

The theoretical uncertainty is estimated by independent  variation of the factorization and renormalization scales by a factor of two up and down, avoiding the extreme combinations (7-point variation) and it is shown by the band in the figures. 

In all predictions, we do not consider effects from multiparton interactions.

\subsection{Comparison to QCD measurements}
\begin{figure}[h!tb]
\begin{center}
\includegraphics[width=0.45\textwidth]{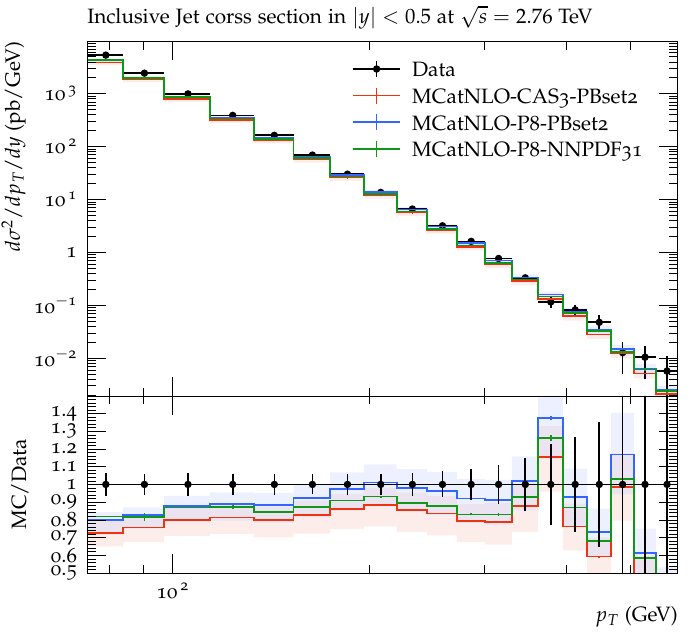}
\includegraphics[width=0.45\textwidth]{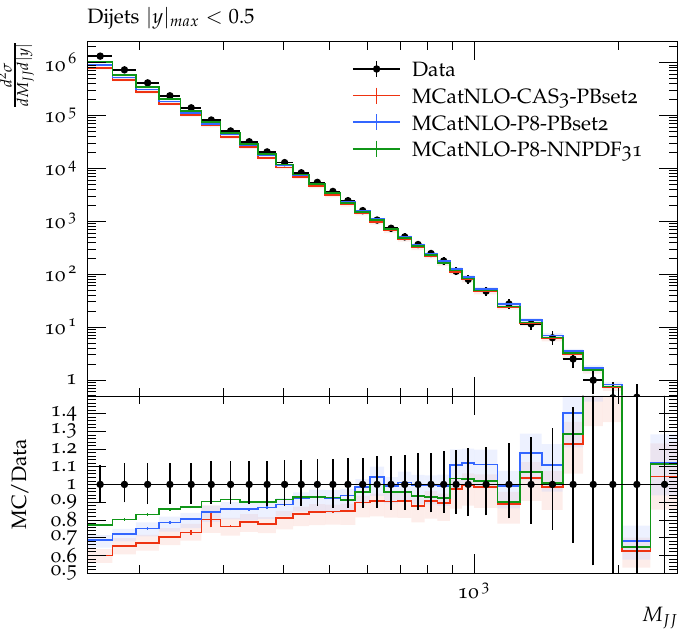}
  \caption{\small Differential cross section as a function of the jet $\pt$ at $\sqrt{s}=2.76$ \TeV~\protect\cite{CMS:2015jdl} (left) and as a function of the dijet mass $M_{JJ}$ at $\sqrt{s}=7$ \TeV~\protect\cite{CMS:2011nlq}  (right) obtained with \protect\CAS\ and \protect\mcatnlo+\pythia 8 using PB-NLO-2018-Set 2. For comparison also shown is  \protect\mcatnlo+\pythia 8 with NNPDF31.}
\label{jets-2.76-7TeV}
\end{center}
\end{figure}

\begin{figure}[h!tb]
\begin{center}
\includegraphics[width=0.45\textwidth]{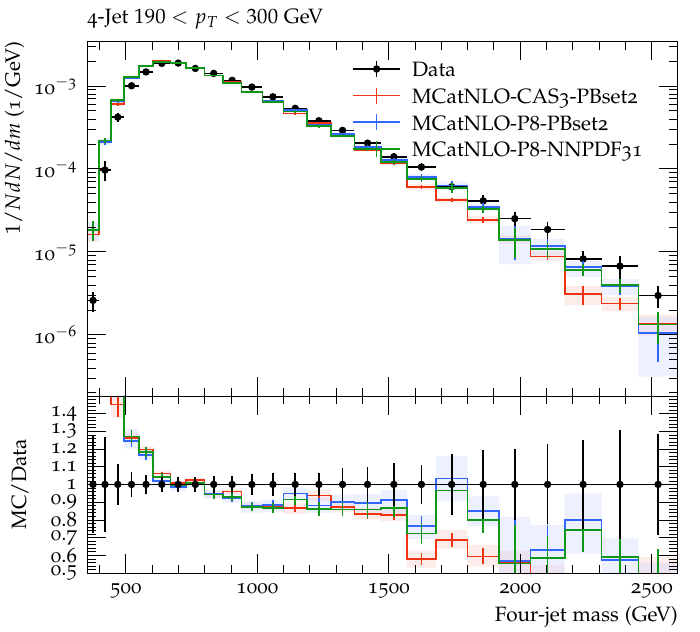}
\includegraphics[width=0.45\textwidth]{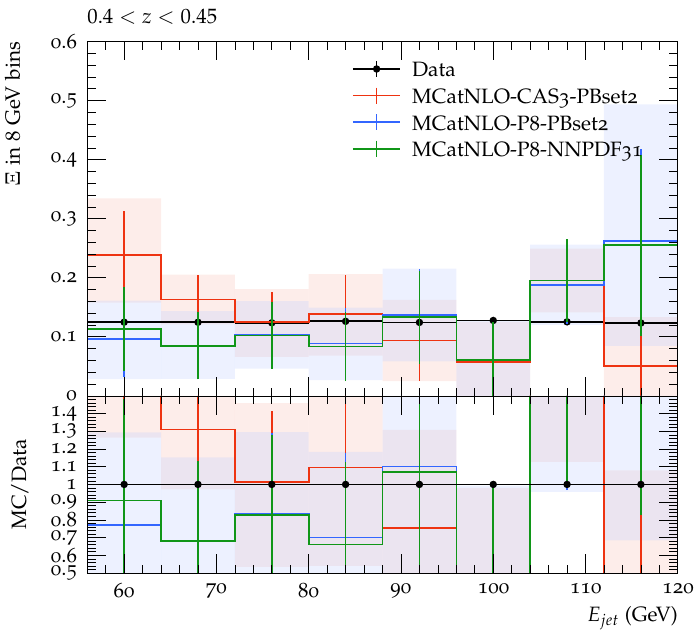}
  \caption{\small Normalized distribution for $x_3$ in high \pt\ three-jet events at $\sqrt{s}=7$ \GeV~\protect\cite{CMS:2015lzt}  (left) and the normalized distribution of energy  of J/$\psi$ mesons inside jets  at $\sqrt{s}=8$ \GeV~\protect\cite{CMS:2019ebt}  (right) obtained with \protect\CAS\ and \protect\mcatnlo+\pythia 8 using PB-NLO-2018-Set 2, for comparison also shown is  \protect\mcatnlo+\pythia 8 with NNPDF31}
\label{jets-7-8TeV}
\end{center}
\end{figure}

\begin{figure}[h!tb]
\begin{center}
\includegraphics[width=0.32\textwidth]{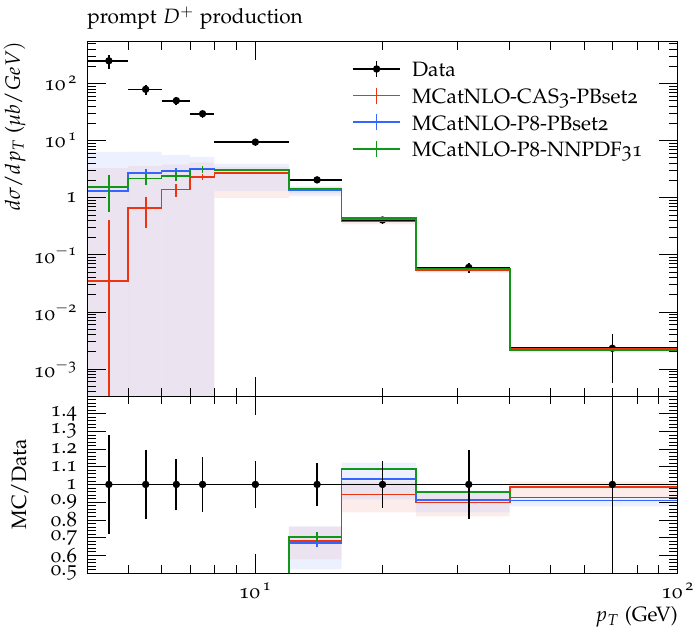}
\includegraphics[width=0.32\textwidth]{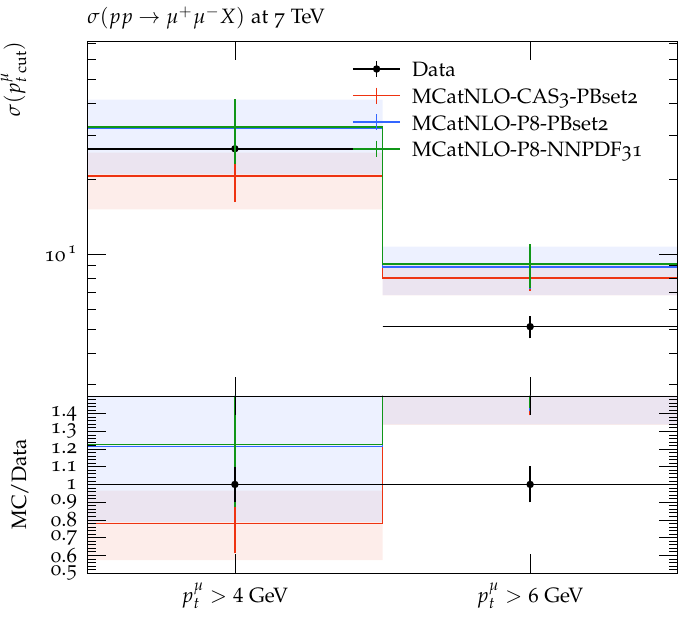}
\includegraphics[width=0.32\textwidth]{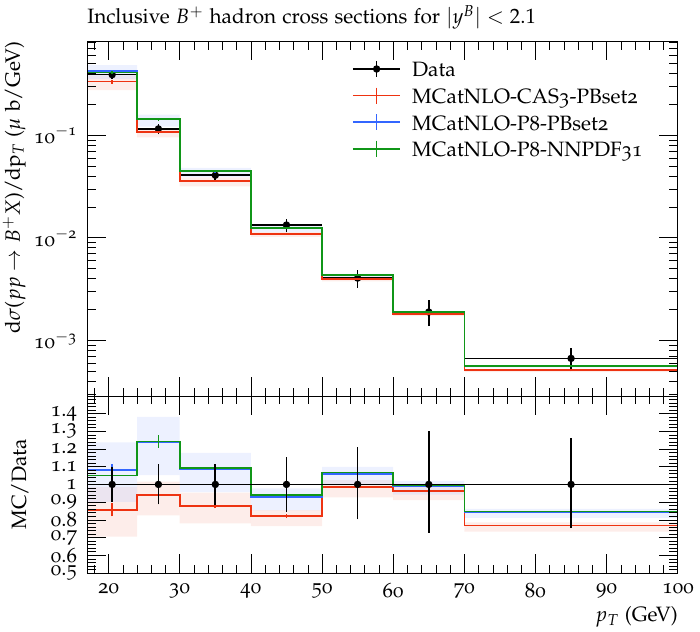}
  \caption{\small Differential cross section as a function of  \pt\ for $D^*$ production at $\sqrt{s}=13$ \TeV~\protect\cite{CMS:2021lab}  (left), the total cross section for $\sigma (pp \to b\bar{b} X \to \mu \mu X^{\prime}$ 
  at $\sqrt{s}=7$ TeV as a function of the \pt\ cut (middle) \protect\cite{CMS:2012xsp}  and the differential cross section for $B^+$ mesons  at $\sqrt{s}=13$ \TeV~\protect\cite{CMS:2016plw}   (right) obtained with \protect\CAS\ and \protect\mcatnlo+\pythia 8 using PB-NLO-2018-Set 2, for comparison also shown is  \protect\mcatnlo+\pythia 8 with NNPDF31.}
\label{HF-7-13TeV}
\end{center}
\end{figure}

\begin{tolerant}{5000}
In Figs.~\ref{jets-2.76-7TeV}--\ref{HF-7-13TeV} we compare measurements from CMS with predictions obtained from \CAS\ and \mcatnlo+\pythia 8 using dijet production at NLO. The aim of this comparison is to check for differences in predictions from \CAS\ and \mcatnlo+\pythia 8 , since both predictions are based on different assumptions for higher order corrections in the form of parton showers. In general we observe a rather good agreement between the two different predictions. 

In Fig.~\ref{jets-2.76-7TeV} inclusive jet cross sections are shown, the calculations give very similar predictions, since inclusive jet and dijet production is largely sensitive to the collinear parton distributions. Also the prediction using NNPDF31 agrees well with the others.
\end{tolerant}
In the following distributions the effect of TMD parton densities and parton shower is investigated.
In Fig.~\ref{jets-7-8TeV}(left) the invariant mass distribution of four-jet events with a \pt\ of the leading jet  of $190 < \pt < 300 $ \GeV\ is shown. The four-jet mass distribution is sensitive to the parton shower, since in the dijet matrix element at NLO at most three jets are produced, and therefore the fourth jet comes from parton shower. 
In Fig.~\ref{jets-7-8TeV}(right) the particle content of jet evolution is investigated by a measurement of the energy of $J/\psi$ mesons inside jets with a transverse momentum of $ \pt > 25$ \GeV\ at $\sqrt{s}=8$ \TeV .  The predictions are in good agreement with the measurements.

Next, heavy flavor production is studied. 
In Fig.~\ref{HF-7-13TeV} measurements of charm and bottom production are shown. Since heavy quarks are not only produced in the hard matrix element, but also during the jet evolution, we use the zero-mass-variable-flavor-number scheme (ZMVFS) for the calculation (which allows charm or bottom quarks to be produced from light parton matrix elements). However, a drawback of ZMVFS is, that the matrix element becomes divergent for small \pt\ (since all partons are treated massless), and therefore a \pt\ cut-off has to be applied. For NLO calculations with subtraction terms (in the \mcatnlo\ frame), we found, that we need to freeze the scale $\mu$ for low $\pt  $. With the procedure described in eq.(\ref{freeze}) we are able to use ZMVFS at NLO down to $\pt = 10 $ \GeV. One can clearly see  the effect of this lower \pt -cut in the distributions of the transverse momentum of heavy flavoured mesons in Fig.~\ref{HF-7-13TeV}. Apart from the low \pt -region, which is affected by the \pt\ cut, the predictions agree rather well with the measurements and only small differences between the different parton shower approaches are observed.

\subsection{Comparison to Onium processes}
Measurements of \jpsi\ production have been performed in \cite{CMS:2011rxs,CMS:2010nis}.  We use the \cascade\ Monte Carlo event generator~\cite{Baranov:2021uol,Jung:2010si,Jung:2001hx} in standalone mode with internally implemented processes based on \kt -factorization with the un-integrated gluon density JH-2013-set1~\cite{Hautmann:2013tba}. Prompt \jpsi\ and $\psi(2S)$  are calculated within the color singlet model, as well as $\chi_c$ production for feed-down to \jpsi . A description of the calculation is given in Ref.~\cite{Baranov:2002cf}. 
\begin{figure}[h!tb]
\begin{center}
\includegraphics[width=0.4\textwidth]{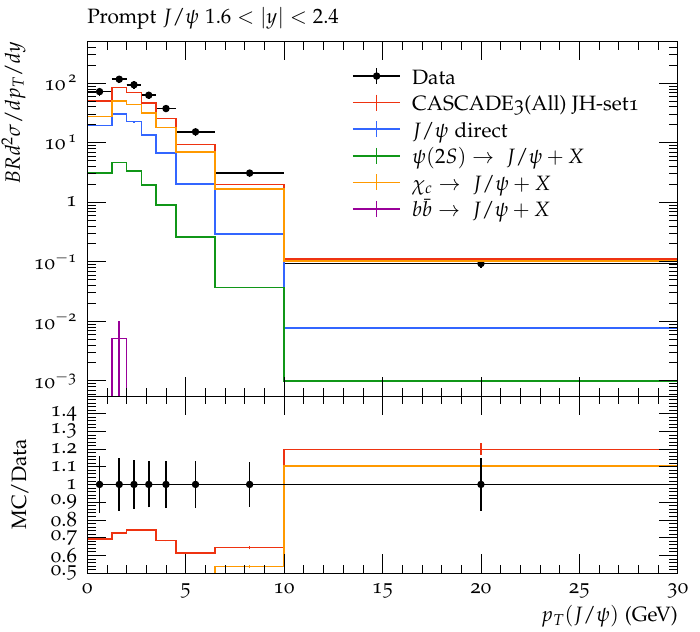}
\includegraphics[width=0.4\textwidth]{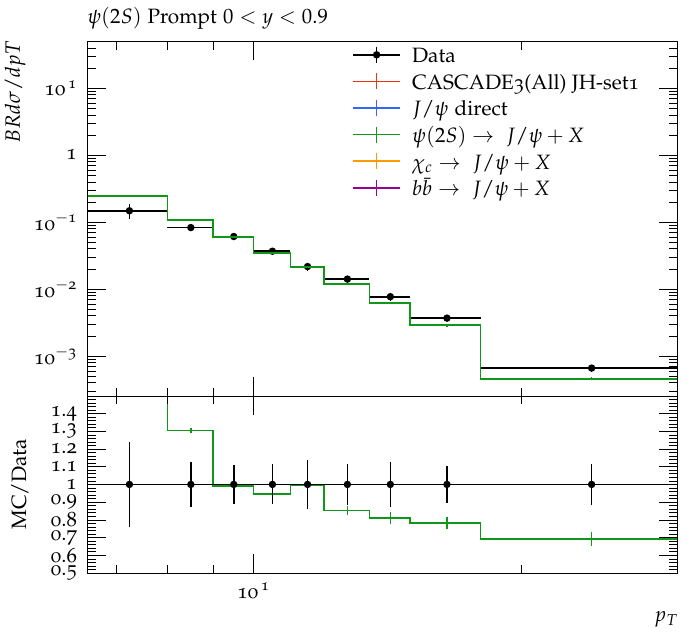}
  \caption{\small Double differential cross section for \jpsi\ production at $\sqrt{s}=7$ \TeV~\protect\cite{CMS:2010nis} (left), and for $\psi(2S)$ production~\protect\cite{CMS:2011rxs} (right) compared with predictions from \cascade 3.
  \label{psi-7TeV}}
\end{center}
\end{figure}
The non-prompt production \jpsi\ is generated using $g^*g^* \to b \bar{b}$ with subsequent decay to \jpsi\ via the hadronization package \pythia 6.
In Fig.~\ref{psi-7TeV}, we show the double differential cross section for prompt \jpsi\ production~\cite{CMS:2010nis} as well as the cross section for prompt $\psi(2S)$ production~\cite{CMS:2011rxs}. A rather good description of the measurements with the predictions based on \kt -factorization is achieved.

\subsection{Comparison to Electroweak processes}
In this section we compare measurements of electroweak processes with predictions from \CAS\ and \protect\mcatnlo+\pythia 8 (with the same settings for parton densities and parton showers as in the previous section) using DY production at NLO.

\begin{figure}[h!tb]
\begin{center}
\includegraphics[width=0.45\textwidth]{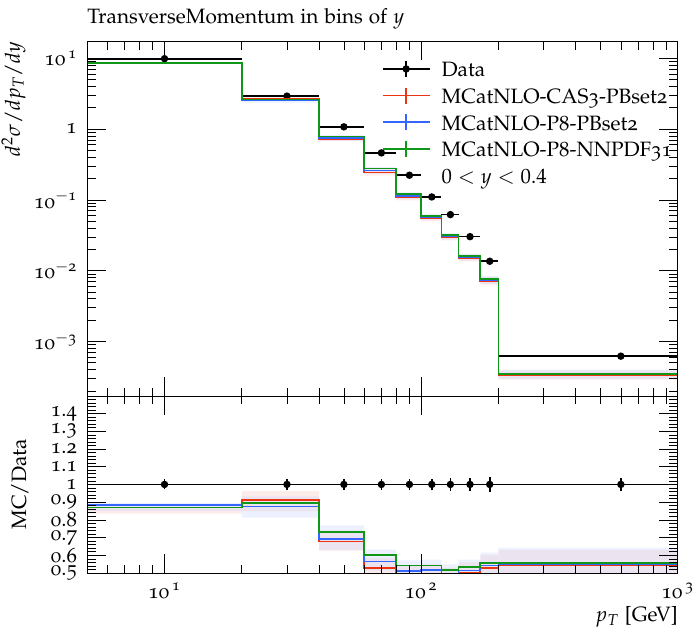}
\includegraphics[width=0.45\textwidth]{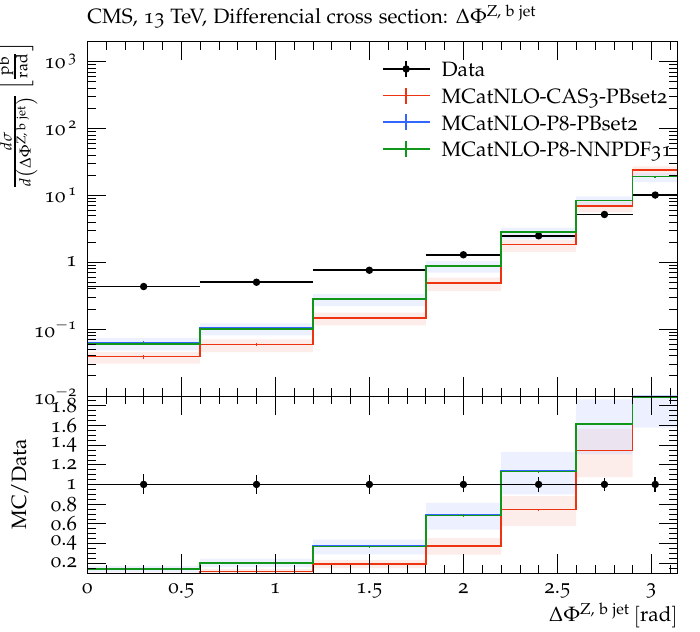}
  \caption{\small Differential cross section as a function of  \pt\ for Z production at $\sqrt{s}=8$ \TeV~\protect\cite{Khachatryan:2015oaa}  (left), and as a function of the azimuthal angle $\Delta \phi$ between a b-jet and the Z boson in Z+b events  at $\sqrt{s}=13$ \TeV~\protect\cite{CMS:2021pcj}   (right) obtained with \protect\CAS\ and \protect\mcatnlo+\pythia 8 using PB-NLO-2018-Set 2. For comparison also shown is  \protect\mcatnlo+\pythia 8 with NNPDF31.}
\label{Z-8-13TeV}
\end{center}
\end{figure}

In Fig.~\ref{Z-8-13TeV}(left) the transverse momentum of Z bosons at $\sqrt{s}=8$ \TeV\ is shown. The calculations  give similar predictions. At larger \pt\ the predictions fall below the measurements, since at larger \pt\ higher order corrections become important. 

In Fig.~\ref{Z-8-13TeV}(right) the cross section as a function of the azimuthal angle $\Delta \phi$ between a b-jet and the Z boson in Z+b events  at $\sqrt{s}=13$ is shown.   Here we compare predictions obtained with inclusive Z boson production at NLO with the measurements and the difference of the predicted cross section to the measurement, especially at small $\Delta \phi$, can be attributed to higher order  hard jet radiation, which is not included in the simulation. Small differences between the parton shower and the TMD predictions can be observed.

\begin{figure}[h!tb]
\begin{center}
\includegraphics[width=0.45\textwidth]{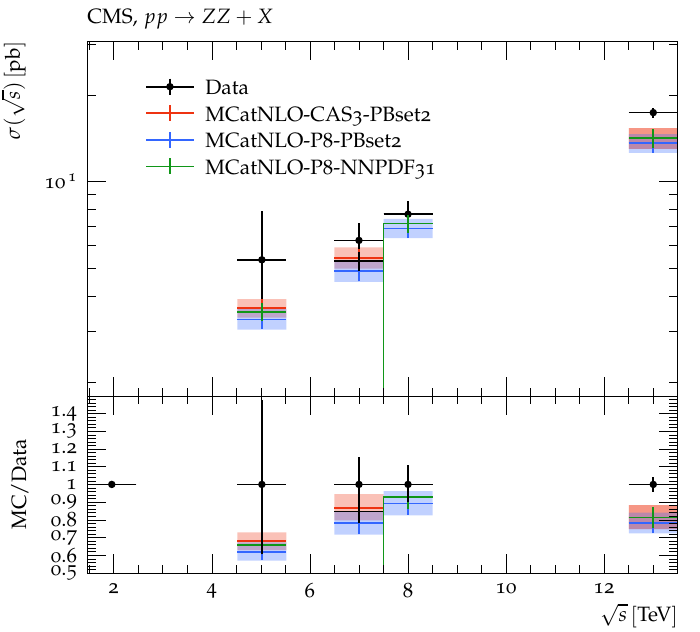}
\includegraphics[width=0.45\textwidth]{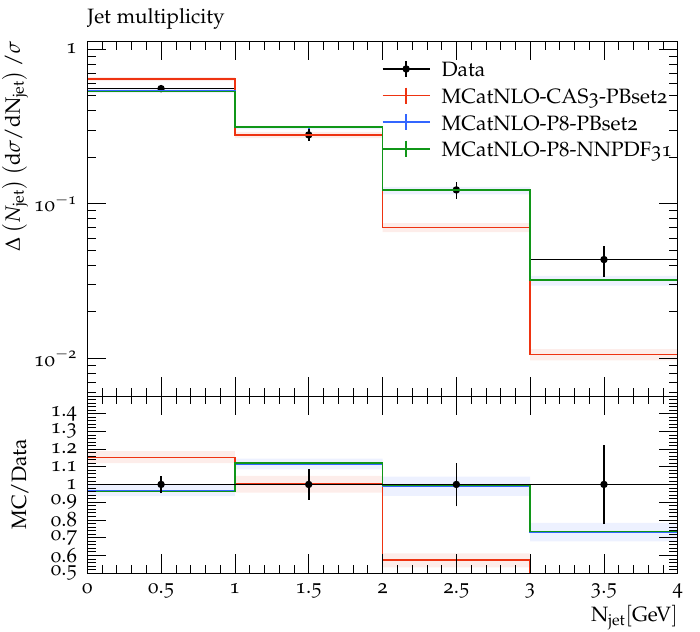}
  \caption{\small  Measurements di-boson production compared with predictions obtained with \protect\CAS\ and \protect\mcatnlo+\pythia 8 using PB-NLO-2018-Set 2. Total cross section for ZZ production as a function of $\sqrt{s}$ (left)~\protect\cite{CMS:2021pqj}, jet multiplicity for WZ production at $\sqrt{s}=13$ \TeV~\protect\cite{CMS:2021icx}   (right). For comparison also shown is  \protect\mcatnlo+\pythia 8 with NNPDF31}
\label{VV}
\end{center}
\end{figure}

In Fig.~\ref{VV}(left) the total cross section of \PZ \PZ\ production as a function of $\sqrt{s}$ as measured in  \cite{CMS:2021pqj}  is compared with predictions obtained from calculations  obtained with a hard process of $\Pp \Pp \to \PZ \PZ +X$ at NLO with \CAS\ and \mcatnlo+\pythia 8.

In Fig.~\ref{VV}(right) the jet multiplicity in WZ events is shown. The hard process is calculated in the mode $\PW\PZ \to l  \nu  l^+ l^-$ with \MCatNLO . Differences coming from the different parton showers can be observed in the jet multiplicity.

\begin{figure}[h!tb]
\begin{center}
\includegraphics[width=0.45\textwidth]{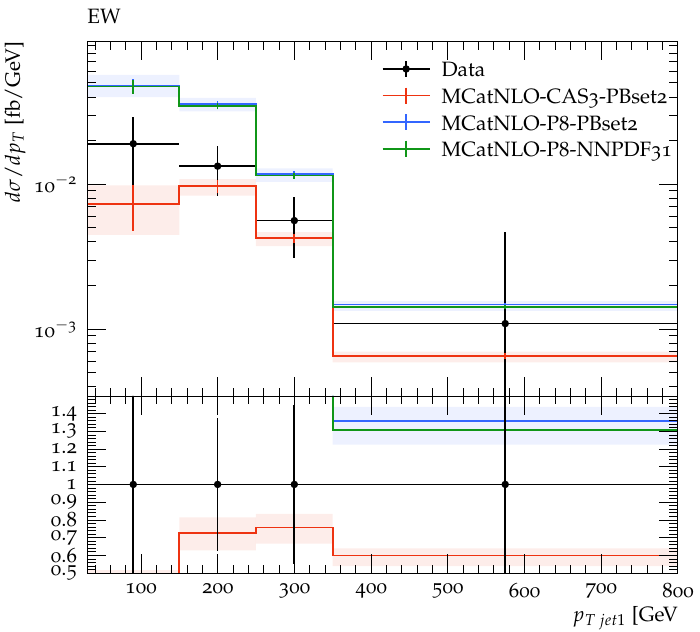}
\includegraphics[width=0.45\textwidth]{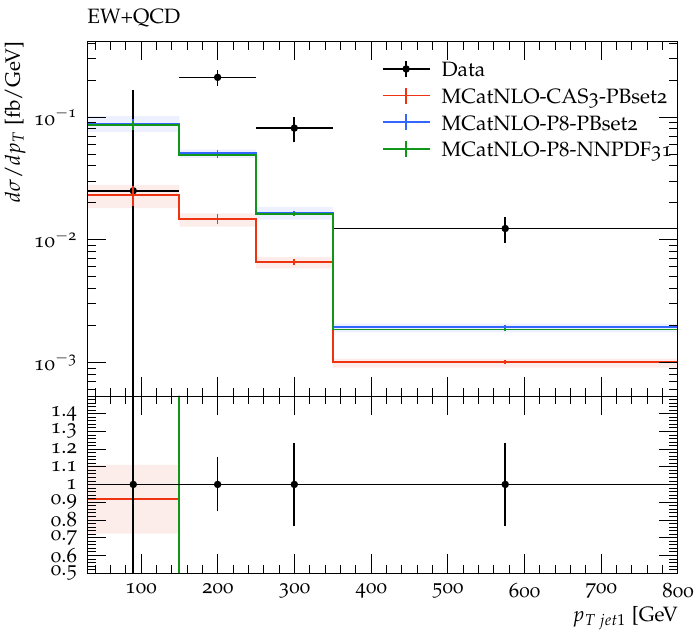}
  \caption{\small Jet \pt\ distribution for events in Z$\gamma$jj  events at $\sqrt{s}=13$ \TeV~\protect\cite{CMS:2021gme}  obtained with \protect\CAS\ and \protect\mcatnlo+\pythia 8 using PB-NLO-2018-Set 2.
  Left: distribution for events in the EW phase space, right: distribution of events in the full EW+QCD phase space.
   }
\label{Zgamma-13TeV}
\end{center}
\end{figure}

In Fig.~\ref{Zgamma-13TeV} the cross section for \PZ$\gamma$jj  at $\sqrt{s}=13$ \TeV~\protect\cite{CMS:2021gme}  as a function of the leading jet \pt\ is shown for events in the EW phase space (left) and in the full EW+QCD phase space (right). For the prediction, the hard process generated  is Z+2jet at NLO, the photon is radiated as part of the final state parton shower, thus representing only the pure QCD phase space. A significant difference between the predictions of \CAS\  and \mcatnlo+\pythia 8 is observed, which can be attributed to the treatment of high \pt\ photon radiation in the parton shower.

\section{Conclusion}
During August 2023 students from different countries joined the {\it Special Remote DESY summer-school} investigating analyses performed by the CMS experiment at the LHC.  Several analyses  in the QCD and electroweak area were implemented  into the Rivet analysis frame for easy comparison with predictions of Monte Carlo event generators.

The analyses were validated by comparing the results of simulations with the ones described in the original publications. 
The results presented here use the newly coded Rivet plugins to perform a comparison of predictions obtained with \CAS\ based Parton Branching transverse momentum dependent  (PB-TMD) parton distributions with the corresponding parton shower and with  \mcatnlo+\pythia 8  applying collinear parton densities and parton showers. The hard processes were generated with  \mcatnlo\ at NLO with the collinear parton density  PB-NLO-2018-Set 2 (and for comparison NNPDF31). It is observed that the difference between predictions obtained with different collinear parton densities is very small, however in some distributions differences are observed between the conventional parton shower in \pythia 8 and the TMD based parton shower (and TMD distributions) in \CAS .

The comparisons presented here are the first systematic comparisons of predictions based on \CAS\ and \mcatnlo+\pythia 8 applying the same collinear parton densities, and therefore contribute to a better understanding of the differences between predictions based on TMD parton densities and conventional parton showers.

The very successful cooperation and collaboration of students with different backgrounds contribute significantly to the Science4Peace idea.

\vskip 1 cm 
\begin{tolerant}{8000}
\noindent 
{\bf Acknowledgments.} 
\noindent 
We are grateful to Andy~Buckley and Christian~G\"utschow from the Rivet team for their support and advice before, during and after the summer-school.
We are also grateful to Markus Seidel for his advice on missing RIVET plugins in CMS as well as for his support in technical details of the analyses. 
Help and support was also received from Achim Geiser on Onium production as well as from Patrick Connor on jet production analyses in CMS.
\end{tolerant} 
\vskip 0.6cm

\bibliographystyle{mybibstyle-new.bst}
\raggedright  
\providecommand{\href}[2]{#2}\begingroup\raggedright\endgroup

\end{document}